\documentclass[10pt]{iopart}
\usepackage[T1]{fontenc} % for copiable characters with accents over them
\usepackage{amstext}
\usepackage{iopams}

\input{glyphtounicode}
\usepackage[
	bookmarks,
	breaklinks,
	colorlinks,
	allcolors=blue,
	urlcolor=blue,
	linktocpage,
	hyperfootnotes,
	bookmarkstype=toc,
	bookmarksopen,
	bookmarksopenlevel=1
]{hyperref}
\hypersetup{pdfpagemode=UseNone}

\usepackage{textcomp}
\usepackage{graphicx}% Include figure files
\usepackage{dcolumn}% Align table columns on decimal point
\usepackage{bm}% bold math
\usepackage[all]{hypcap}

\usepackage{xcolor}
\definecolor{grey}{rgb}{0.5,0.5,0.5}
\definecolor{lightgrey}{rgb}{0.9,0.9,0.9}
\usepackage{iopams}

\begin{document}

\title[Phase diffusion in trapped-atom interferometers]{Phase diffusion in trapped-atom interferometers}

\author{Valentin Ivannikov$^{1-4}$ and Andrei I. Sidorov$^3$}

\address{$^1$ New York University Shanghai, Shanghai, China}
\address{$^2$ State Key Laboratory of Precision Spectroscopy, School of Physics and Materials Science, East China Normal University, Shanghai, China}
\address{$^3$ Centre for Quantum and Optical Science, Swinburne University of Technology, Melbourne, Australia}
\address{$^4$ Universidade de S\~ao Paulo, Instituto de F\'isica de S\~ao Carlos, S\~ao Carlos, Brazil}
\ead{valentin@nyu.edu}
\vspace{10pt}
\begin{indented}
\item[]\today
\end{indented}

\begin{abstract}
We evaluate the performance and phase diffusion of trapped $^{87}$Rb atoms in an atom-chip sensor with Ramsey interferometry and Hahn's spin echo in the time and phase domains. We trace out how the phase uncertainty of interference fringes grows with time. The phase-domain spin echo enables us to attain many-second-long phase diffusion with a low-cost local oscillator that otherwise seems unrealistic to obtain with such an oscillator. In the Ramsey experiment we record interference fringes with contrast decaying in $12$~s, and with a frequency uncertainty of $80$~mHz corresponding to the dephasing time of $2.8$~s. A clear distinction is drawn between the decoherence of the atomic ensemble, and the dephasing originating from the local oscillator. Spin echo cancels most of the perturbations affecting the Ramsey experiments, and leaves the residual phase noise of only $19$~mHz mostly attributed to the local oscillator frequency instability, yielding the increased coherence time of $11.9$~s which coincides with the contrast decay time in the Ramsey sequence. A number of perturbation sources leading to homogeneous and inhomogeneous dephasing is discussed. Our atom-chip sensor is useful in probing fundamental interactions, atomtronics, microcantilever, and resonant cavity profiling \textit{in~situ}.
\end{abstract}

\ioptwocol

\section{Introduction}

Subtleties of thermal binary spin mixtures, exhibiting spin waves, are on a surge of exploration in Ramsey \cite{Deutsch2010a,KleineBuning2011,Bernon2013} and beyond-Ramsey protocols \cite{Solaro2016,Yudin2016} the simplest of which is spin echo \cite{Hahn1950}.
Compact atomic sensors based on these protocols are suitable for sensing different types of fields and are long-awaited in portable, airborne, marine, Earth- and space-based metrology\footnote{Cold-atom research in microgravity \cite{VanZoest2010}, e.g., the NASA's Cold Atom Laboratory developed by the Jet Propulsion Laboratory.}. In emerging fields of atomic physics such as atomtronics, micromechanics with ultracold atoms, photonics with atoms confined in a hollow-core photonic crystal fiber \cite{Epple2014}, control over quantum states has become even more challenging \cite{Treutlein2012}.
These novel $\mu$m-scale systems pose non-trivial calibration and characterization problems \cite{Shatil2017}.
Even more demanding is quantum metrology where one can measure evanescent shifts due to cold atomic collisions \cite{Harber2002}, black-body radiation \cite{Rosenbusch2007short,Middelmann2012short}, probe many-body states in quantum metamaterials \cite{Bernien2017short,Lanyon2017short}, or meter fundamental interactions \cite{Greene2017short,Moses2017,Fletcher2017short}.
Such applications require advanced decoherence treatment, hence, accurate ways of coherence profiling. Yet, full decoupling from decohering bath remains unrealized and further studies on the dynamic decoupling \cite{Uhrig2007,Sagi2010,Wimperis1994}, especially in the regime of spin locking \cite{Naydenov2011}, can be anticipated.

\begin{figure}[!t]
\centering{\includegraphics[width=1\columnwidth]{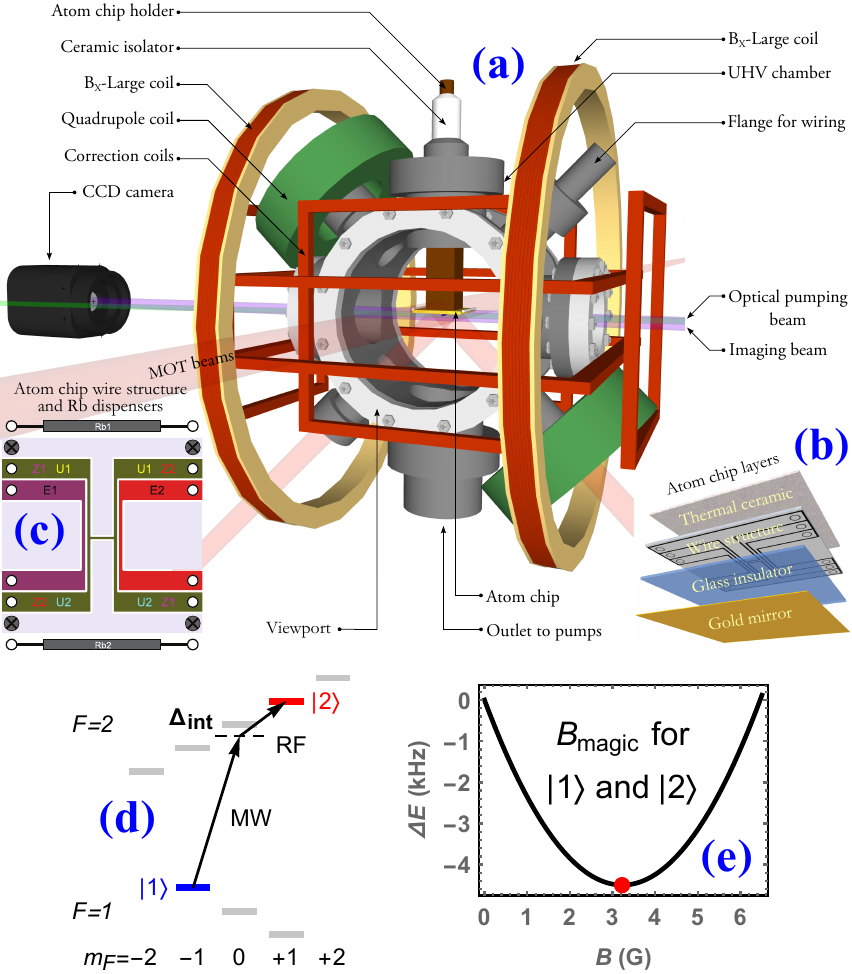}}
\caption[Physical geometry of UHV chamber, coils and atom chip]{Apparatus with the vacuum chamber (a) encompassing the atom chip (b,c), coils, and the laser beams. The atomic cloud is hovering underneath the atom chip (a). Energy levels of the $^{87}$Rb clock states (d), and the ``magic'' field marked with the red dot (e). $\Delta_{\text{int}}$ is the detuning from $\left|2,0\right\rangle$, $\Delta E$ is the energy difference between $\left|1\right\rangle$ and $\left|2\right\rangle$.}
\label{fig:chamber}
\end{figure}

Microtraps of ultracold atoms on atom chips \cite{Folman2002,Fortagh2007short,AtomChips2011} have been applied as compact sensors to asymmetric potentials \cite{Hall2007}, near-surface imaging \cite{Bohi2010} and compact atomic clocks \cite{Szmuk2015}.
However, on-chip interferometers are very attractive for compact, transportable and sensitive metrology.
Our contribution is concentrated on the realization of an ultracold atom interferometer on an atom chip (Fig.~\ref{fig:chamber}), in which we pay special attention to phase diffusion.
In our study two atom interferometry regimes are used.
The atom clock regime \cite{Nicholson2015short} implies the classic Ramsey two pulses and the longest possible interrogation time \cite{Ramsey1949} to generate the error signal for correcting the time counting.
The atomic sensor regime compensates for undesired inhomogeneous dephasing during atomic phase evolution by employing the spin-echo rephasing scheme to purify the dependence of its interference on the measured pertubation.
In both regimes decoherence is the key problem to preserve the quantum state \cite{Kuhr2005,Szmuk2015}, and it can be significantly alleviated with sophisticated techniques \cite{Windpassinger2008,Wimperis1994}.

Decoherence that eventually limits the precision of quantum sensors, can be detected through interferometric contrast decay \cite{Deutsch2010a,KleineBuning2011,Solaro2016} or through the growth of phase uncertainty \cite{Egorov2011}. To improve the understanding of the deleterious phase diffusion, we present an alternative way of characterizing coherence properties via the dynamics of the phase diffusion, rather than the visibility decay. Our interferometry signal is extracted from an ensemble average. Inhomogeneous dephasings in trapped atoms, originating from collisionally induced frequency shifts \cite{Harber2002,Rosenbusch2009}, are density dependent and are different for the two velocity classes in an ensemble \cite{Deutsch2010a}. Effects of inhomogeneous dephasing can be substantially reduced by using the Hahn's spin echo technique or by rephasing the result of the binary spin collisions through the identical spin rotation effect (ISRE) \cite{Lhuillier1982,Deutsch2010a}. Irreversible homogeneous dephasing \cite{Kuhr2005} originates from either frequency instability of the local oscillator, or from magnetic field noise.

\section{Experimental sequence}
\label{sec:expapparatus}

\begin{figure*}[ht]
\includegraphics[width=1\textwidth]{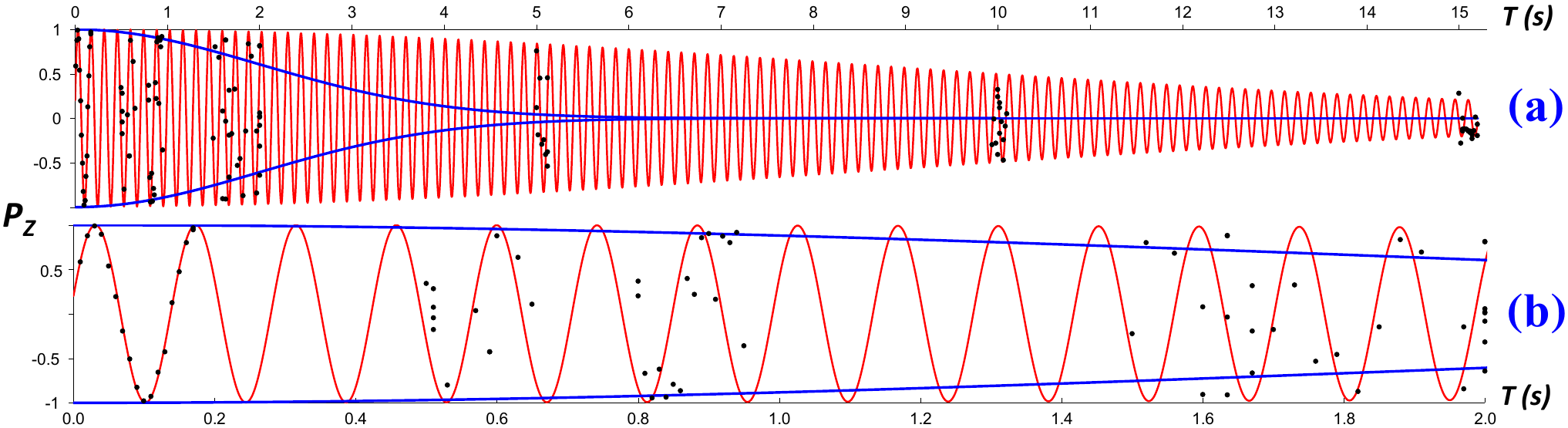}
\caption[Ramsey $15$~s]{Time-domain Ramsey interference of trapped atoms. (a) Long time decay of the atomic superposition (black dots) with the contrast envelope reconstructed from atomic coherence time $\tau = 12$~s with function $e^{-(T/\tau)^2}\cos(43.1T+0.319)$ (red curve). (b) Scattering of data at the short evolution time originating from the microwave (MW) frequency instability and atom number variations with the contrast decay fitted with $e^{-(T/\tau_d)^2}$ (blue curve) and the dephasing time $\tau_d = 2.83$~s.}
\label{fig:ramsey12s}
\end{figure*}

We realize the trapped atom interferometer in two hyperfine states $\left|1\right\rangle \equiv \left|F=1,m_F=-1\right\rangle$ and $\left|2\right\rangle \equiv \left|2,+1\right\rangle$ of $^{87}$Rb atoms in a magnetic microtrap on an atom chip \cite{Hall2007,Anderson2009short,Egorov2011} (Fig.~\ref{fig:chamber}).
We use laser cooling and evaporative cooling of $50,000$ atoms in state $\left|1\right\rangle$ down to the temperature of $(250\pm10)$~nK at the mean atomic density $\bar{n}=n_0/\sqrt{8} = 0.31\times 10^{12}$~cm$^{-3}$, where $n_0$ is the peak atomic density. Our sequence of laser cooling, magnetic trapping, and evaporative cooling is described in Refs.~\cite{Anderson2009short,Egorov2011,Ivannikov2013thesis}.

At the ``magic'' value of $3.229$~G the $1^\text{st}$-order differential Zeeman shift is nullified, and the coherent superposition of states $\left|1\right\rangle$ and $\left|2\right\rangle$ is largely insensitive to the magnetic field noise and can maintain coherence for up to $60$~s owing to the identical spin rotation effect (ISRE) \cite{Lhuillier1982,Deutsch2010a}. We employ the two-photon transition with the $6.831$~GHz MW field and $3.25$~MHz radio frequency (RF) field to prepare the coherent superposition of states $\left|1\right\rangle$ and $\left|2\right\rangle$.
The Zeeman splitting of states $\left|2\right\rangle$ and $\left|2,0\right\rangle$ is $2.258$~MHz.
We prevent populating the intermediate $\left|2,0\right\rangle$ state by using detuning 1~MHz from the intermediate  state.
Microwave radiation is generated by an Agilent E8257D generator, synchronized either with an internal OCXO or with a Stanford Research Systems (SRS) FS725 rubidium frequency standard.
The rubidium atoms are trapped in both states in a cigar-shaped magnetic trap with frequencies $\{97.6, 97.6, 11.96\}$~Hz.
RF field is produced by the SRS DS345 generator and it is synchronized with the MW synthesizer to avoid frequency drift between the pulses.
The Rabi frequencies of MW and RF couplings are determined to be $7.2$~kHz and $67.8$~kHz, respectively, from the Rabi oscillations, and the two-photon resonant Rabi oscillations at $39.8$~Hz yield the $\pi/2$-pulse duration of $1$~ms.

We simultaneously detect the state populations $N_1$ and $N_2$ by adiabatic passage $\left|1\right\rangle \rightarrow \left|2,-2\right\rangle$, the application of Stern-Gerlach separation, and absorption imaging with saturation correction \cite{Reinaudi2007short}.
In the reported experiments the observable is the normalized population difference $P_z = (N_1 - N_2)/(N_1 + N_2)$ that carries the spin-ensemble phase accumulated during evolution. Absorption imaging destroys the cloud, hence the entire cooling sequence cycles with $T_{cycle} = 83$~s to collect other $P_z$ data points from Ramsey or spin echo.

In the Ramsey sequence the first MW/RF $\pi/2$-pulse prepares a $50$:$50$ coherent superposition of the two states, and synchronizes an atomic field with the electromagnetic interrogating field. Immediately after the pulse, the ensemble phase starts evolving as $\phi = \Delta T + \varphi_{col}$, where $\Delta = \omega_{21} - \omega_{MW} - \omega_{RF}$, and $\varphi_{col}$ is the density dependent collisional shift causing inhomogeneous dephasing \cite{Kuhr2005}. The second $\pi/2$-pulse with a variable MW phase $\phi_0$ projects the accumulated phase onto the $Z$-axis of the Bloch sphere to give $P_z$.
Interferometric fringes are obtained in $P_z$ by either varying $T$ (Fig.~\ref{fig:ramsey12s}), or the phase of the second $\pi/2$ pulse by changing the MW phase $\phi_0$. In Ramsey interferometry we mix two oscillators: an atomic superposition of states $\left|1\right\rangle$ and $\left|2\right\rangle$, and the two-photon electromagnetic field. Our atomic oscillator has better phase stability than the MW/RF oscillator. We shift our attention from the time-domain Ramsey to spin echo in the phase domain, and show that the spin echo removes inhomogeneous dephasing leaving the depasing due to the local oscillator frequency instability.
We use two methods to analyze the phase-domain data.

\subsection{Method~1: $\sigma$-extraction from phase fringe}

It is used when data contains a full MW-phase scan producing a complete interferometric fringe at the fixed evolution time $T$. We assume a Gaussian distribution of the frequency uncertainty \cite{Kuhr2005,Rosenbusch2009} with the mean value $\Delta$ and standard deviation $v_\sigma$. We fit the phase data with $P_z = V e^{-\frac{1}{2} v_\sigma^2 T^2} \cos(\Delta T + \phi_0)$, where $\Delta$ is the fitted detuning and $V$ is the fitted contrast. The sought phase uncertainty standard deviation $\sigma$ is calculated as $\sigma = v_\sigma T$. The errors are calculated as $\pm SE = \pm \sigma N_{pts}^{-1/2}$, where $N_{pts}$ is the number of points.

\subsection{Method~2: $\sigma$ from zero-crossing signal}

Here we collect data with fixed $T$ and the MW phase $\phi_0$ at which the $P_z$ mean is zero (Fig.~\ref{fig:zerocrossing2}). MW phase fluctuations lead to the observed $P_z$ fluctuations and we evaluate phase values as $\phi\left(P_z\right) = \arccos(P_z/V) - \phi_0$. We again assume a Gaussian distribution of phase fluctuations and from the set of phase data evaluate the phase standard deviation $\sigma$ and the MW frequency standard deviation $v_\sigma = \sigma / T$. Contrast $V$ is found from a full phase scan of a fringe. The dephasing time of the MW oscillator is defined as $\tau_{d} = \sqrt{2}/v_\sigma$.

To enhance the data accuracy, the dependence of $P_z$ on the total number of atoms $N$ should be taken into account before obtaining $v_\sigma$ from the fit (Fig.~\ref{fig:zerocrossing2}).
Phase shift is a linear function of a mean density and thus of an atom number $N$ of trapped thermal atoms.
In Method~2 the $P_z\left(N\right)$ dependence is assumed linear for small variations of $P_z$ \cite{Egorov2011,Szmuk2015}.
By subtracting the fitted function from the measured points of $P_z\left(T\right)$ or $P_z\left(\phi\right)$ we remove the irrelevant contribution from them.
We call this number correction ($N$-corr) and it is not applicable in Method~1.

\section{Ramsey interference decay}
\label{sec:Decay}

\begin{figure}[t]
\centering{\includegraphics[width=0.96\columnwidth]{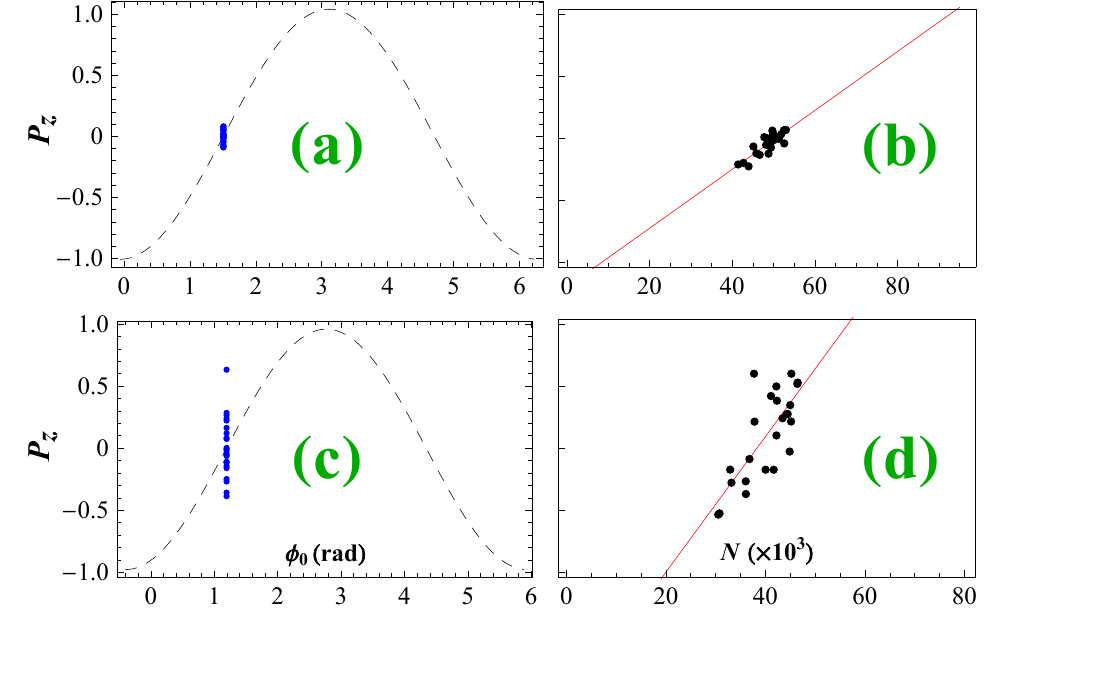}}
\caption[Zero-crossing Ramsey interferometry and atom number fluctuations]{Zero-crossing Ramsey interferometry in the phase domain with $N$-corr~\textbf{(a,c)}, and in the N-domain without $N$-corr~\textbf{(b,d)} with $5\times 10^4$ of non-condensed atoms at $250$~nK. An OCXO installed in the MW-generator clocks the two-photon field. Plots \textbf{(a,c)} correspond to measurements at evolution times $T = \{100, 500\}$~ms with $N$-corr. Data points (blue) are corrected for the $P_z(N)$ fluctuations having approximately linear dependence (red in \textbf{(b,d)}). The $P_z(\phi)$ cosine (grey dashed) is fitted to show zero-crossing $P_z = 0$. The fitted $\sigma$ values after $N$-corr are $\{47.6, 239\}$~mrad for \textbf{(a)} and \textbf{(c)}, respectively, resulting in a fitted $\sigma = 479 T$~mrad. Without $N$-corr $\sigma = \{89.3, 375\}$~mrad and $\sigma = 755 T$~mrad.}
\label{fig:zerocrossing2}
\end{figure}

Figure~\ref{fig:ramsey12s} shows drastic competition between ISRE and the phase diffusion from the two oscillators (which are the atomic state superposition and the two-photon interrogating field) in Ramsey interference. ISRE \cite{Lhuillier1982,Deutsch2010a} is driven by binary elastic collisions in the trapped ensemble and is characterized by $\bar{n}$.

First, Fig.~\ref{fig:ramsey12s}(a) demonstrates long-lasting atomic interference with a slowly decaying contrast.
We fit this decay with $e^{-(T/\tau)^2}$ to extract the atomic coherence time $\tau = 12$~s.
Collisional dephasing $4 \pi \hbar (a_{11} - a_{22}) \bar{n}m^{-1}$ = $2 \pi \times 0.12$~Hz is dominated by the spin exchange \cite{Lhuillier1982}, since the ISRE rephasing rate $\omega_{ex} = 4 \pi \hbar a_{12} \bar{n}m^{-1} = 2 \pi \times 2.4$~Hz \cite{Deutsch2010a}, where $a_{11}$ and $a_{22}$ above are the intrastate scattering lengths, $a_{12}$ is the interstate scattering length in $^{87}$Rb atoms \cite{Egorov2011}, and $m$ is the atomic mass.
The slow decay of interference contrast likely occurs due to magnetic field noise perturbing the transition through the second order Zeeman effect.

Second, on the short time scale (Fig.~\ref{fig:ramsey12s}(b)) we observe the deviation of the data points from the oscillations reconstructed from the frequency drift and the instability of our MW generator Agilent E8257D. In this measurement it is clocked by a Rb frequency standard (SRS FS725).
The RF generator at $3.25$~MHz is synchronized with the same reference and does not contribute to the observed phase instability.
According to the FS725 specification the fractional frequency instability is $\Delta f/f = 10^{-11}$ at sampling times from $1$ to $10$~s.
This nominal frequency instability amounts to $v_{\sigma, th}$ = $2 \pi \times 68$~mHz and the dephasing time of $\tau_{d,th} = 3.3$~s.
From the stochastic distribution of data points at three intervals of the evolution time $\{0-0.17, 0.5-0.65, 0.8-0.95\}$~s of Fig.~\ref{fig:ramsey12s}(b) we evaluate the frequency uncertainty $v_\sigma = 2 \pi \times 80$~mHz and measure the dephasing time $\tau_{d} = 2.8$~s. which is close to the theoretical value of $3.3$~s.

We also carried out measurements of phase diffusion at four evolution times using the zero-crossing method (Fig.~\ref{fig:zerocrossing2}).
In this sequence our microwave oscillator was clocked by the internal oven-controlled crystal oscillator (OCXO) which has the fractional frequency instability of $\Delta f/f = 5 \times 10^{-12}$ at sampling times $1-10$~s according to the Agilent specification.
Experimental points in Fig.~\ref{fig:zerocrossing2}(a,c) were corrected for shot-to-shot atom number variations.
Different measurements of $P_z$ at the same evolution time and the same MW phase reveal data point scattering around the linear trend line in the atom number space (Fig.~\ref{fig:zerocrossing2}(b,d)).
Without $N$-corr the phase uncertainties are $\sigma = \{89.3, 375\}$~mrad and the frequency uncertainty standard deviation was $v_\sigma = 2 \pi \times 120$~mHz.
By fitting $P_z(N)$ linearly we obtain a slope of $2.38\times 10^{-5}$ for Fig.~\ref{fig:zerocrossing2}(b) and $5.47\times 10^{-5}$ for Fig.~\ref{fig:zerocrossing2}(d).
The fitted linear trend is subtracted from $P_z(N)$ before extracting $\sigma$ values.
Two measurements of the phase spread at $T = \{100, 500\}$~ms result in $\sigma = \{47.6, 239\}$~mrad, respectively.
Overall, we made four measurements of phase uncertainty at $T = \{0.1, 0.2, 0.4, 0.5\}$~s (Fig.~\ref{fig:dephasingexp}) to extract the frequency spread of $v_\sigma = 2 \pi \times 80$~mHz, with $N$-corr.

\begin{figure}[b]
\centering{\includegraphics[width=0.8\columnwidth]{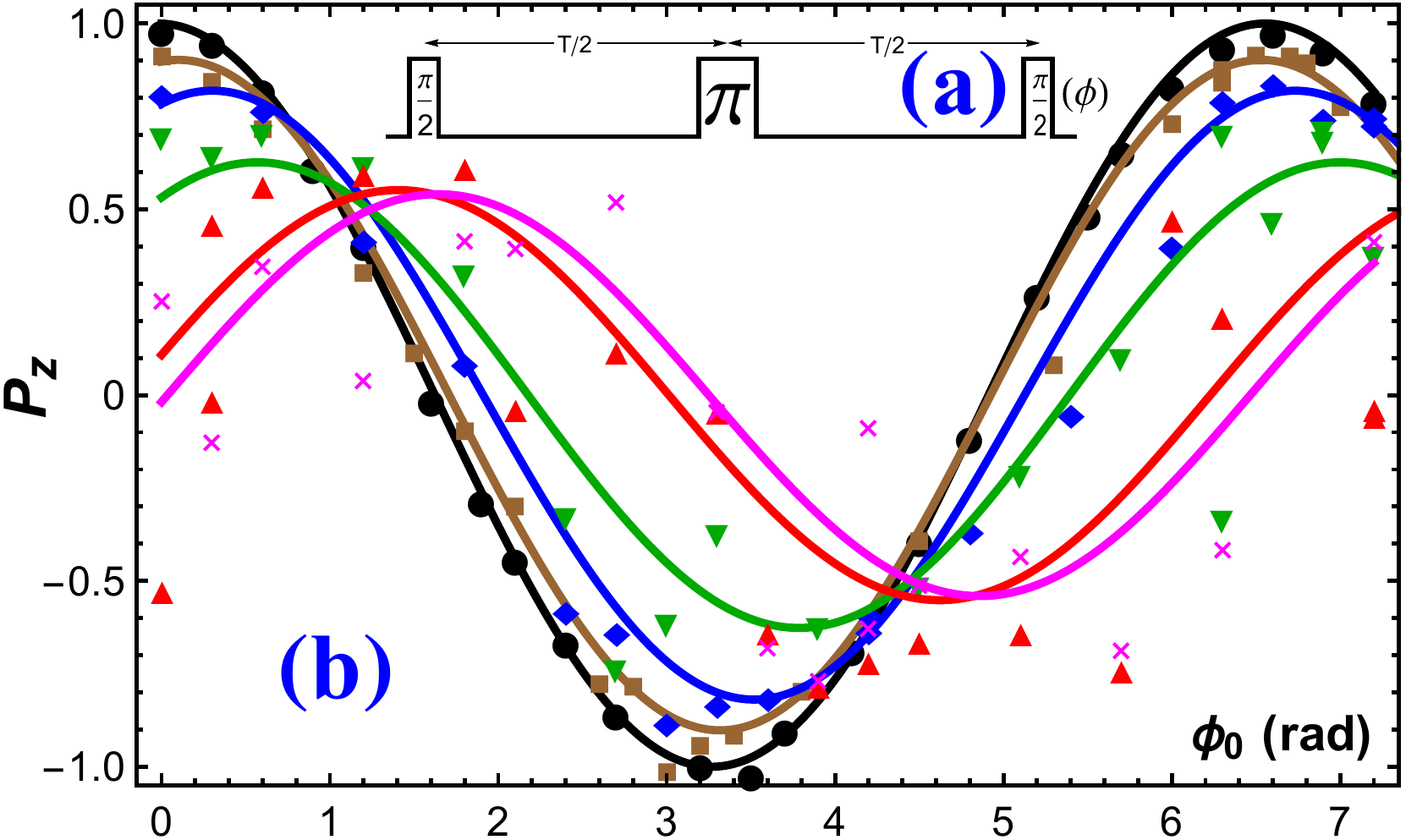}}
\caption[Phase-domain spin echo with OCXO]{Phase-domain spin echo measurements and the fits with $T=\{0.2, 1, 2, 4, 6, 8\}$~s corresponding to %$\{\markerone, \markersix, \markerfive, \markertwo, \markerthree, \markerfour \}$
\{black, brown, blue, green, red, magenta\}. $5\times 10^4$ thermal atoms in $B_{\text{magic}}$ were trapped at a temperature of $250$~nK. Spin echo drastically reduces phase uncertainty as compared to the Ramsey experiment: at $T = 8$~s the fringe still clearly follows a cosine.}%
\label{fig:phaseechoexp}%
\end{figure}%

\section{Phase diffusion in spin echo}
\label{sec:Spin-echo}

In the spin echo sequence the $\pi$-pulse is applied at time $T/2$ (Fig.~\ref{fig:phaseechoexp}(a)) to reverse inhomogeneous dephasing \cite{Kuhr2005,Ivannikov2013thesis}.
By the time of the second $\pi/2$ pulse collisional shift and other inhomogeneities are almost rephased, and $P_z$ would show a maximal contrast, unless other dephasing effects are present such as the frequency instability of the local oscillator or residual magnetic noise of $\omega_{21}$.
The ISRE effect competes with the spin echo rephasing at higher densities. It can deteriorate its performance and even reverse its effect \cite{Solaro2016}.

Our MW generator is clocked by an internal OCXO and we employ Method~1 to measure phase diffusion and the contrast decay for evolution times up to $T=8$~s (Fig.~\ref{fig:phaseechoexp}(b)), an order of magnitude longer, than for the Ramsey sequence.
At the shortest $T = 0.1$~s we record the minimal phase diffusion of $19$~mHz, and the interference contrast of $0.99$.
At the longest $T=8$~s the phase standard deviation increases up to $0.75$~rad and the contrast reduces to $0.55$.

\begin{figure}[t]%
\centering{\includegraphics[width=0.96\columnwidth]{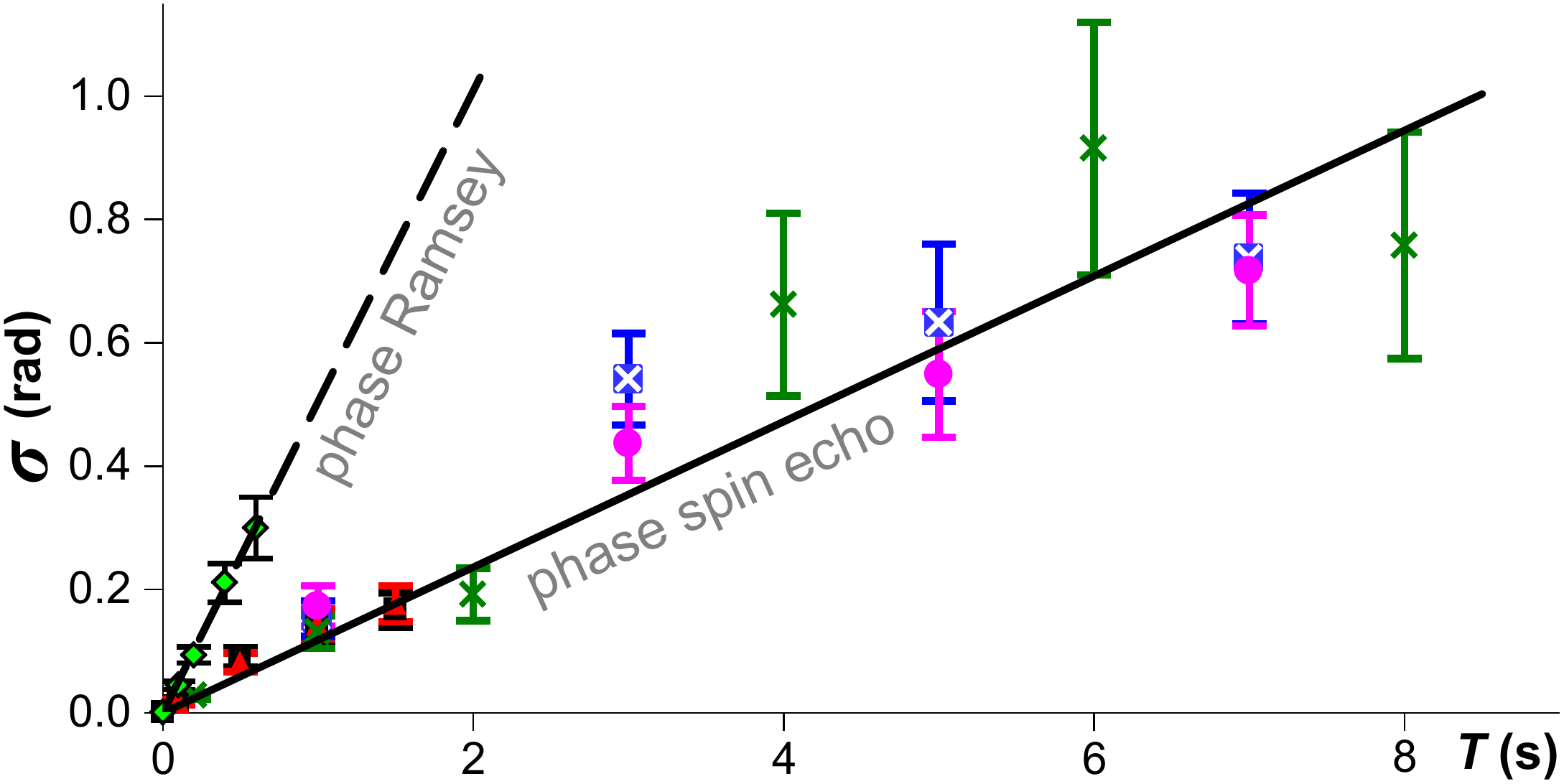}}%
\caption[Phase-domain Ramsey and spin echo phase diffusion with OCXO]{Phase diffusion with phase-domain Ramsey (dashed line) and spin echo (solid line) clocked by the OCXO. The points at $T \leq 2$~s are measured with Method~2, beyond $2$~s according to Method~1. Color coding indicates independent measurement sessions. The fit of Ramsey data gives $v_\sigma = 2 \pi \times 80$~mHz and that of spin echo gives $2 \pi \times 19$~mHz.}%
\label{fig:dephasingexp}%
\end{figure}%

In Fig.~\ref{fig:dephasingexp} the growth of phase uncertainty $\sigma(T)$ with the evolution time in two interferometric sequences is presented.
The application of spin echo greatly reduces inhomogeneous broadening by a factor of $4$ down to $19$~mHz providing the dephasing time of $11.9$~s, very close to the independently measured $12$-s Ramsey contrast decay (Fig~\ref{fig:ramsey12s}(a)).

\section{Discussion}
\label{sec:Discussion}

The time-domain Ramsey interference (Fig.~\ref{fig:ramsey12s}) provides us two values of decoherence time: the contrast decay time $\tau = 12$~s and the dephasing time $\tau_d = 2.83$~s evaluated from the recorded frequency uncertainty as $80$~mHz.
The zero-crossing measurements with the more stable OCXO (Fig.~\ref{fig:dephasingexp}) give us the same value of $80$~mHz.
In the spin echo sequence we measure the dephasing rate of $19$~mHz, and the dephasing time $11.9$~s.

Several noise sources affect phase stability in trapped-atom interferometers.
These include detection noise, magnetic field noise, temperature fluctuations, atom number variations, collisional broadening, and local oscillator instability \cite{Kuhr2005,Rosenbusch2009,Szmuk2015}. States $\left|2\right\rangle$ and $\left|1\right\rangle$ have different two-body collisional decays that may lead to a lower visibility \cite{Ivannikov2014,Ivannikov2017}.
Projection noise $\left(\Delta f/f\right)_{N} = (2\pi T f V\sqrt{N})^{-1}$ does not limit us, according to our estimates.
Ambient magnetic noise is coupled to the two-photon transition via the $2^\text{nd}$ order differential Zeeman shift.
There are magnetic field fluctuations of $5$~mG in the atomic trap location measured by a fluxgate magnetometer .
We use Eq.~(18) in \cite{Szmuk2015} to estimate the frequency uncertainty of only $8$~mHz at the ``magic'' field due to the magnetic noise of $5$~mG.

Shot-to-shot temperature variations lead to atom number variations, an associated collisional shift and broadening of the transition frequency and the variable range of magnetic fields confining the atomic cloud.
The collisional broadening at our $\bar{n}$ is compensated to the great extent by the ISRE rephasing in the both Ramsey and spin echo sequences.
For our conditions spin echo assists in cancelling phase inhomogeneities and does not destructively compete with the ISRE effect \cite{Solaro2016}.
However, shot-to-shot atom number variations lead to the shot-to-shot fluctuations of the transition frequency and to the frequency noise in our $P_z$ data.
In our measurements of phase diffusion in Ramsey and spin echo sequences we make $N$-correction  to the recorded phase.
In the data of Fig.~\ref{fig:zerocrossing2} the atom number uncertainty is $4\% (\pm$ 2,000 atoms).
In the zero-crossing Ramsey interferometry the $N$-corr slope is $5.47\times 10^{-5}$ leading to the associated phase uncertainty of $0.11$~rad and the frequency noise of $35$~mHz.
In order to estimate independently the contribution of temperature fluctuations to phase diffusion in our setup we use the fractional frequency Allan deviation value at the magic field in Fig.~11 in \cite{Szmuk2015}, scale it up with our temperature uncertainty ($10$~nK) to obtain the same frequency uncertainty of $34$~mHz.

An independent source of dephasing in our experiment is the instability of MW frequency locked to an OCXO with the specified fractional instability $\Delta f/f = 5 \times 10^{-12}$ and the corresponding frequency uncertainty of $34$~mHz.
This specified uncertainty is larger than our recorded frequency instability of $19$~mHz in the spin echo measurement so we assumed that our actual instability of the local oscillator is smaller by a factor of two ($17$~mHz).
Assuming independent influence of magnetic noise ($8$~mHz), atom number uncertainty due to temperature fluctuations ($35$~mHz)  and the local oscillator noise ($17$~mHz)and adding all these contributions, we estimate the standard deviation of frequency instability of our trapped atom interferometer as $40$~mHz which is below our measured value of $80$~mHz in the Ramsey sequence.
Spin echo reverses the collisional shift due to the atom number variations to remove the temperature fluctuations contribution.
The two remaining factors (magnetic noise and local oscillator instability) provide us with an estimate of a frequency uncertainty of $19$~mHz.

\section{Conclusions}
\label{sec:Conclusions}

Coherent superposition of magnetically trapped $^{87}$Rb atoms in $\left|1\right\rangle$ and $\left|2\right\rangle$ states at the ``magic'' magnetic field can be a stable reference to lock MW field and serve as a compact portable atomic clock on a chip \cite{Szmuk2015}.
Previous studies on how long coherence in a such atomic oscillator can survive \cite{Deutsch2010a,KleineBuning2011,Szmuk2015,Solaro2016} were focused on recording the decay of interference contrast. Paper \cite{KleineBuning2011} reports the contrast decay time of $21$~s but states that fringes wash out after some seconds due to frequency noise.
Our study here has focused on monitoring how phase uncertainty grows with time in Ramsey and spin echo sequences to measure a dephasing time of $11.9$~s which was limited by frequency instability of our commercial MW generator and can be significantly extended by using more stable sources. 
We demonstrated that spin echo can substantially suppress inhomogeneous dephasing and provide dephasing rates by a factor of four smaller than in the conventional Ramsey interferometry.

The reported trapped atom interferometer has broad applicability in quantum information and metrology \cite{Jing2018arXiv}, atom clocks, atomtronics, studies of fundamental interactions, or the characterization of Rydberg atoms and their plasmas via the motional Stark effect \cite{Kaiser2017short,Levinton1989short}.
It is indispensable in accurate tuning of a micromechanical device, e.g., microcantilever, or a resonant cavity interacting with an ultracold cloud or Bose-Einstein condensate.
A compact trapped-atom clock with the contrast decay time of $58$~s \cite{Deutsch2010a} already demonstrated a fractional instability of $5.8 \times 10^{-13}$ with a $1$-s integration and can show a stability of $10^{-15}$ a day \cite{Szmuk2015}. 
The ability to maintain coherence for tens of seconds can be used to store qubit information in a trapped-atom ensemble and serve as quantum memory.

\ack
V.I. acknowledges support from S\~ao Paulo Research Foundation (FAPESP) via grant 2016/23874-4.

\section*{References}

\bibliography{references2}

\providecommand{\newblock}{}
\begin{thebibliography}{10}
\expandafter\ifx\csname url\endcsname\relax
  \def\url#1{{\tt #1}}\fi
\expandafter\ifx\csname urlprefix\endcsname\relax\def\urlprefix{URL }\fi
\providecommand{\eprint}[2][]{\url{#2}}

\bibitem{Deutsch2010a}
Deutsch C, Ram\'{\i}rez-Mart\'{\i}nez F, Lacro\^{u}te C, Reinhard F, Schneider
  T, Fuchs J~N, Pi\'{e}chon F, Lalo\"{e} F, Reichel J and Rosenbusch P 2010
  {\em Phys. Rev. Lett.\/} {\bf 105} 020401

\bibitem{KleineBuning2011}
{Kleine B\"{u}ning} G, Will J, Ertmer W, Rasel E, Arlt J, Klempt C,
  Ram\'{\i}rez-Mart\'{\i}nez F, Pi\'{e}chon F and Rosenbusch P 2011 {\em Phys.
  Rev. Lett.\/} {\bf 106}(3) 240801

\bibitem{Bernon2013}
Bernon S, Hattermann H, Bothner D, Knufinke M, Weiss P, Jessen F, Cano D,
  Kemmler M, Kleiner R, Koelle D and Fort\'{a}gh J 2013 {\em Nat. Commun.\/}
  {\bf 4} 2380

\bibitem{Solaro2016}
Solaro C, Bonnin A, Combes F, Lopez M, Alauze X, Fuchs J~N, Pi\'echon F and
  Pereira Dos~Santos F 2016 {\em Phys. Rev. Lett.\/} {\bf 117}(16) 163003

\bibitem{Yudin2016}
Yudin V~I, Taichenachev A~V, Basalaev M~Y and Zanon-Willette T 2016 {\em Phys.
  Rev. A\/} {\bf 94} 052505

\bibitem{Hahn1950}
Hahn E~L 1950 {\em Phys. Rev.\/} {\bf 80} 580

\bibitem{VanZoest2010}
van Zoest T, Gaaloul N, Singh Y, Ahlers H, Herr W, Seidel S~T, Ertmer W, Rasel
  E, Eckart M, Kajari E, Arnold S, Nandi G, Schleich W~P, Walser R, Vogel A,
  Sengstock K, Bongs K, Lewoczko-Adamczyk W, Schiemangk M, Schuldt T, Peters A,
  K\"{o}nemann T, M\"{u}ntinga H, L\"{a}mmerzahl C, Dittus H, Steinmetz T,
  H\"{a}nsch T~W and Reichel J 2010 {\em Science (N.Y.)\/} {\bf 328} 1540 ISSN
  1095-9203 \urlprefix\url{http://www.ncbi.nlm.nih.gov/pubmed/20558713}

\bibitem{Epple2014}
Epple G, Kleinbach K~S, Euser T~G, Joly N~Y, Pfau T, Russell P~S~J and L{\"o}w
  R 2014 {\em Nat. Commun.\/} {\bf 5} 4132

\bibitem{Treutlein2012}
Treutlein P 2012 {\em Science\/} {\bf 335} 1584

\bibitem{Shatil2017}
Shatil N~R, Homer M~E, Picco L, Martin P~G and Payton O~D 2017 {\em Appl. Phys.
  Lett.\/} {\bf 110} 223101

\bibitem{Harber2002}
Harber D~M, Lewandowski H~J, McGuirk J~M and Cornell E~A 2002 {\em Phys. Rev.
  A\/} {\bf 66} 053616

\bibitem{Rosenbusch2007short}
Rosenbusch P, Zhang S and Clairon A 2007 {Blackbody radiation shift in primary
  frequency standards} {\em IEEE IFCS-EFTF\/} p 1060 ISBN 978-1-4244-0646-3
  ISSN 1075-6787

\bibitem{Middelmann2012short}
Middelmann T, Falke S, Lisdat C and Sterr U 2012 {\em Phys. Rev. Lett.\/} {\bf
  109} 263004

\bibitem{Bernien2017short}
Bernien H, Schwartz S, Keesling A, Levine H, Omran A, Pichler H, Choi S, Zibrov
  A~S, Endres M, Greiner M, Vuleti{\'c} V and Lukin M~D 2017 {\em Nature\/}
  {\bf 551} 579--584

\bibitem{Lanyon2017short}
Lanyon B~P, Maier C, Holzapfel M, Baumgratz T, Hempel C, Jurcevic P, Dhand I,
  Buyskikh A~S, Daley A~J, Cramer M, Plenio M~B, Blatt R and Roos C~F 2017 {\em
  Nat. Phys.\/} {\bf 13} 1158--1162

\bibitem{Greene2017short}
Greene C~H, Giannakeas P and P\'erez-R\'{\i}os J 2017 {\em Rev. Mod. Phys.\/}
  {\bf 89}(3) 035006

\bibitem{Moses2017}
Moses S~A, Covey J~P, Miecnikowski M~T, Jin D~S and Ye J 2017 {\em Nat.
  Phys.\/} {\bf 13} 13

\bibitem{Fletcher2017short}
Fletcher R~J, Lopes R, Man J, Navon N, Smith R~P, Zwierlein M~W and Hadzibabic
  Z 2017 {\em Science\/} {\bf 355} 377

\bibitem{Uhrig2007}
Uhrig G~S 2007 {\em Phys. Rev. Lett.\/} {\bf 98} 100504

\bibitem{Sagi2010}
Sagi Y, Almog I and Davidson N 2010 {\em Phys. Rev. Lett.\/} {\bf 105} 053201

\bibitem{Wimperis1994}
Wimperis S 1994 {\em J. Magn. Reson., Ser A\/} {\bf 109} 221

\bibitem{Naydenov2011}
Naydenov B, Dolde F, Hall L, Shin C, Fedder H, Hollenberg L, Jelezko F and
  Wrachtrup J 2011 {\em Phys. Rev. B\/} {\bf 83} 081201

\bibitem{Folman2002}
Folman R, Kr\"{u}ger P, Schmiedmayer J, Denschlag J and Henkel C 2002 {\em
  Advances In Atomic, Molecular, and Optical Physics\/} {\bf 48} 263

\bibitem{Fortagh2007short}
Fort\'{a}gh J and Zimmermann C 2007 {\em Rev. Mod. Phys.\/} {\bf 79} 235

\bibitem{AtomChips2011}
Reichel J and Vuletic V 2011 {\em {Atom Chips}\/} (WILEY-VCH Verlag)

\bibitem{Hall2007}
Hall B~V, Whitlock S, Anderson R, Hannaford P and Sidorov A~I 2007 {\em Phys.
  Rev. Lett.\/} {\bf 98} 030402

\bibitem{Bohi2010}
B\"{o}hi P, Riedel M~F, H\"{a}nsch T~W and Treutlein P 2010 {\em Appl. Phys.
  Lett.\/} {\bf 97} 051101

\bibitem{Szmuk2015}
Szmuk R, Dugrain V, Maineult W, Reichel J and Rosenbusch P 2015 {\em Phys. Rev.
  A\/} {\bf 92} 012106

\bibitem{Nicholson2015short}
Nicholson T~L, Campbell S~L, Hutson R~B, Marti G~E, Bloom B~J, McNally R~L,
  Zhang W, Barrett M~D, Safronova M~S, Strouse G~F, Tew W~L and Ye J 2015 {\em
  Nat. Commun.\/} {\bf 6} 6896

\bibitem{Ramsey1949}
Ramsey N~F 1949 {\em Phys. Rev.\/} {\bf 76} 996

\bibitem{Kuhr2005}
Kuhr S, Alt W, Schrader D, Dotsenko I, Miroshnychenko Y, Rauschenbeutel A and
  Meschede D 2005 {\em Phys. Rev. A\/} {\bf 72}(2) 023406

\bibitem{Windpassinger2008}
Windpassinger P~J, Oblak D, Petrov P~G, Kubasik M, Saffman M, Alzar C~L~G,
  Appel J, M\"uller J~H, Kj\ae{}rgaard N and Polzik E~S 2008 {\em Phys. Rev.
  Lett.\/} {\bf 100}(10) 103601

\bibitem{Egorov2011}
Egorov M, Anderson R~P, Ivannikov V, Opanchuk B, Drummond P~D, Hall B~V and
  Sidorov A~I 2011 {\em Phys. Rev. A\/} {\bf 84} 021605

\bibitem{Rosenbusch2009}
Rosenbusch P 2009 {\em Appl. Phys. B\/} {\bf 95} 227

\bibitem{Lhuillier1982}
Lhuillier C~C and Lalo\"{e} F 1982 {\em J. Phys. (Paris)\/} {\bf 43} 197

\bibitem{Anderson2009short}
Anderson R~P, Ticknor C, Sidorov A~I and Hall B~V 2009 {\em Phys. Rev. A\/}
  {\bf 80} 023603

\bibitem{Ivannikov2013thesis}
Ivannikov V 2013 {\em {Analysis of a trapped atom clock with losses}\/} {PhD
  Thesis} Swinburne University of Technology
  \urlprefix\url{http://www.swinburne.edu.au/engineering/caous/theses.htm}

\bibitem{Reinaudi2007short}
Reinaudi G, Lahaye T, Wang Z and Gu\'{e}ry-Odelin D 2007 {\em Opt. Lett.\/}
  {\bf 32} 3143

\bibitem{Ivannikov2014}
Ivannikov V 2014 {\em Phys. Rev. A\/} {\bf 89} 023615

\bibitem{Ivannikov2017}
Ivannikov V 2017 {\em Phys. Rev. A\/} {\bf 95} 033621

\bibitem{Jing2018arXiv}
Jing Y, Fadel M, Ivannikov V and Byrnes T 2012  (\textit{Preprint}
  \eprint{1808.10679})

\bibitem{Kaiser2017short}
Kaiser M, Grimmel J, Torralbo-Campo L, Mack M, Karlewski F, Jessen F, Schopohl
  N and Fort\'agh J 2017 {\em Phys. Rev. A\/} {\bf 96}(4) 043401

\bibitem{Levinton1989short}
Levinton F~M, Fonck R~J, Gammel G~M, Kaita R, Kugel H~W, Powell E~T and Roberts
  D~W 1989 {\em Phys. Rev. Lett.\/} {\bf 63}(19) 2060

\end{thebibliography}

\end{document}